\documentclass[a4paper,11pt]{article}
\usepackage[utf8]{inputenc}

\usepackage{geometry}
\usepackage{setspace}
\usepackage{authblk}

\usepackage{array, threeparttable, booktabs,caption}
\usepackage{times}
\usepackage{enumitem}
\usepackage{hyperref}

\hypersetup{colorlinks,allcolors=black}
\usepackage{tabularx}
\usepackage{graphicx}
\usepackage{adjustbox}
\newcommand{\RNum}[1]{\uppercase\expandafter{\romannumeral #1\relax}}
\newcommand{\beginsupplement}{%
        \setcounter{section}{0}
        \renewcommand{\thesection}{S\arabic{section}}
        \setcounter{table}{0}
        \renewcommand{\thetable}{S\arabic{table}}%
        \setcounter{figure}{0}
        \renewcommand{\thefigure}{S\arabic{figure}}%
     }

\begin{document}

\title{\textbf{The academic Great Gatsby Curve}}
\author[1]{Ye Sun}
\author[1,2,3]{Fabio Caccioli}
\author[4]{Xiancheng Li}
\author[5,1,*]{Giacomo Livan}

\affil[1]{Department of Computer Science, University College London, 66-72 Gower Street, London WC1E 6EA, United Kingdom}
\affil[2]{London School of Economics and Political Science, Systemic Risk Centre, London WC2A 2AE, United Kingdom}
\affil[3]{London Mathematical Laboratory, United Kingdom}
\affil[4]{School of Business and Management, Queen Mary University of London, Mile End Road, London E1 4NS, United Kingdom}
\affil[5]{Dipartimento di Fisica, Universit\`a degli Studi di Pavia, via Bassi 6, 27100 Pavia, Italy}

\affil[*]{To whom correspondence and requests for materials should be addressed. E-mail: giacomo.livan@unipv.it}
\date{}

\maketitle

\begin{abstract}
The Great Gatsby Curve measures the relationship between income inequality and intergenerational income persistence. By utilizing genealogical data of over 245,000 mentor-mentee pairs and their academic publications from 22 different disciplines, this study demonstrates that an academic Great Gatsby Curve exists as well, in the form of a positive correlation between academic impact inequality and the persistence of impact across academic generations. We also provide a detailed breakdown of academic persistence, showing that the correlation between the impact of mentors and that of their mentees has increased over time, indicating an overall decrease in academic intergenerational mobility. We analyze such persistence across a variety of dimensions, including mentorship types, gender, and institutional prestige.
\end{abstract}

\section*{Introduction}
Intergenerational income mobility, which measures the extent to which income is passed on from one generation to the next, reflects the degree of openness in a society. Lower levels of mobility imply higher class persistence, meaning that an individual's economic status is largely determined by their family background~\cite{chetty2014land,xie2022trends}. In recent years, the issues of declining mobility~\cite{song2020long} and rising inequality~\cite{saez2016wealth}, as well as their interrelationship~\cite{corak2013income}, have attracted considerable attention from scholars and policymakers. The empirically observed positive correlation between income inequality and intergenerational income persistence~\cite{andrews2009more,krueger2012rise} is often  referred to as the Great Gatsby Curve in the literature, based on the 1925 novel by F. Scott Fitzgerald exploring --- among others --- the theme of class persistence. Such a relationship has important implications for understanding the mechanisms of social mobility in different contexts, and the potential policy levers to enhance it.

A number of recent studies have investigated concepts that represent the academic equivalents of inequality and intergenerational persistence. The former refers to the uneven distribution of opportunity~\cite{graves2022inequality} and academic impact~\cite{nielsen2021global,sun2023ranking}, which --- in spite of its multifaceted nature --- is usually operationalized in terms of the volume of citations accrued by publications over time~\cite{fortunato2018science}. The latter is instead quantified by the influence that a mentor's status may have on their prot\'eg\'es' academic success~\cite{malmgren2010role,sekara2018chaperone,li2019early, ma2020mentorship}. 

In this paper we seek to determine whether an ``academic Great Gatsby Curve'' exists, i.e., whether academic inequality and intergenerational persistence are positively correlated. In line with the above studies, we quantify academic inequality as the concentration of impact in a population of authors, as measured by the Gini coefficient of the distribution of citations. We operationalize academic intergenerational persistence as the correlation between the academic impact of mentors and that of their prot\'eg\'es, mirroring the association between parents' and children's economic well-being. In fact, a mentor can sometimes be seen as a mentee's ``academic parent'', as reflected by the German terms for supervisor, Doktorvater or Doktormutter, literally meaning doctoral father or mother. However, unlike the transmission of economic welfare from parents to children, the inheritance between mentors and mentees mainly involves the transfer of research skills and experience, knowledge of the field~\cite{lienard2018intellectual} and professional networks~\cite{sekara2018chaperone,li2019early,ma2020mentorship}. We expect a high level of persistence across academic generations to be associated with unequal opportunities in academia, which we seek to detect as inequality in the distribution of citations across authors in a discipline. 

We already documented a positive relationship between academic impact inequality and lack of mobility in impact rankings in a previous study~\cite{sun2023ranking}. In that case, the notion of mobility we considered was related to different moments in an author's career. Here, instead, we are interested in mobility across academic generations, and therefore in comparing an author's academic status with that of their mentors. This is the closest academic equivalent to intergenerational mobility as considered in the Social Sciences.

In the following, we analyze genealogical data on more than 300,000 academics who published nearly 10 million papers in 22 disciplines from 2000 to 2013 (See \textit{Methods}), examining temporal trends of academic persistence between mentors and their mentees, and comparing such trends across different mentorship types, different mentor-mentee gender combinations and different tiers of institutional prestige. Finally, we document the existence of an academic Great Gatsby Curve, namely the positive relationship between academic impact inequality and academic intergenerational persistence.

\section*{Results}
\begin{figure*}[ht!]
\centering
    \includegraphics[width=16cm]{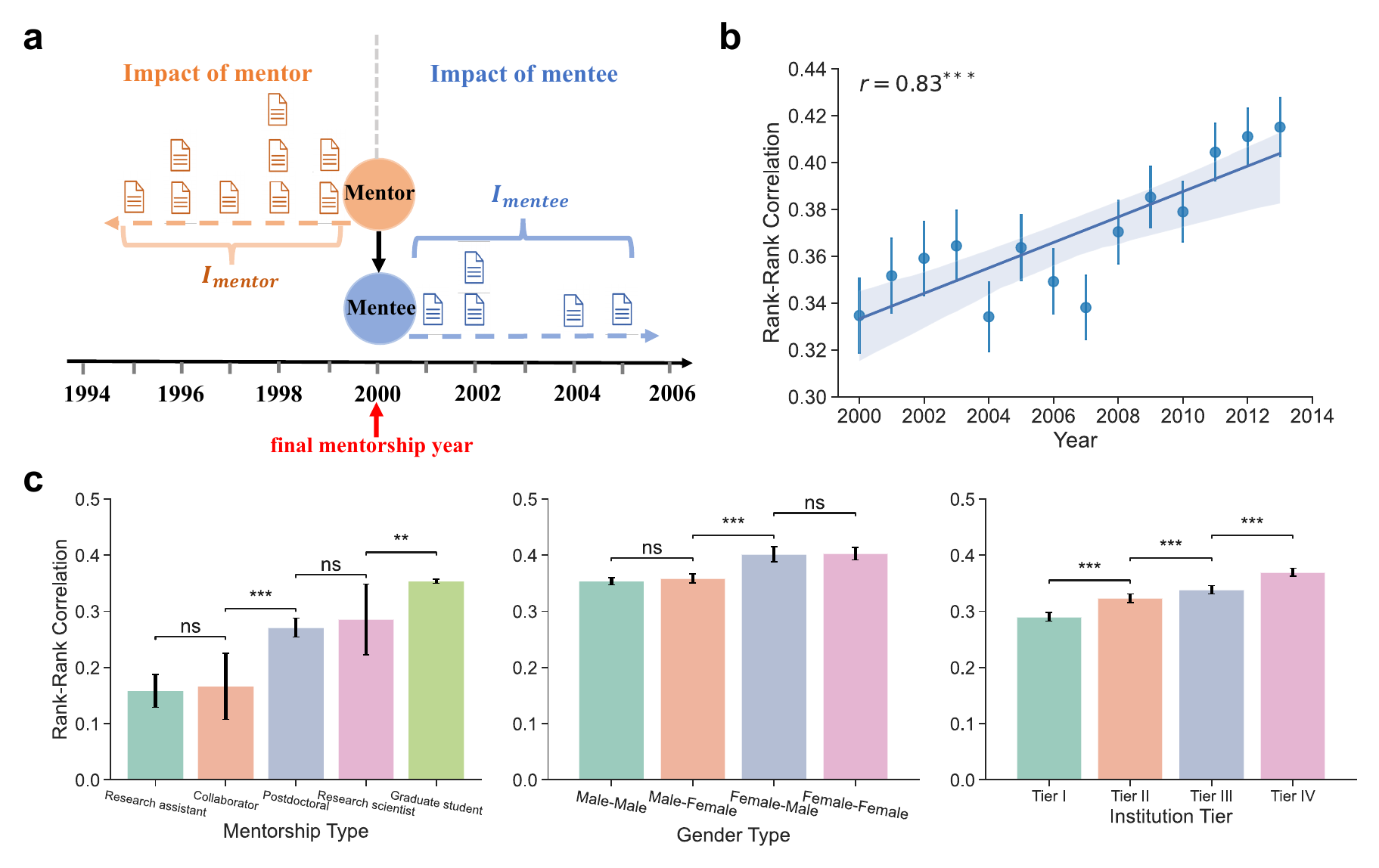}
    \caption{\textbf{Persistence of academic impact between mentors and mentees.} (\textbf{a}) Simple illustration of our assumptions to measure the scientific impact of mentors and mentees within a 5-year time window before and after the final mentorship year. The impact of a mentor/mentee over a period of time is calculated as the total number of citations received by their papers (within 5 years after publication) published over that period. (\textbf{b}) Persistence of academic impact between mentors and their mentees is on the rise. Here, impact persistence is measured as the Spearman's rank correlation coefficient between the positions of mentors and mentees in the impact rankings of their discipline, using data on cohorts of mentor-mentee pairs with the same final mentorship year. The solid line and the shaded area represent the regression line (with annotated Pearson’s $r$ and $p$ values) and the $95\%$ confidence level interval, respectively. (\textbf{c}) Comparison of impact persistence across different mentorship types, mentor-mentee gender combinations and tiers of institutional prestige (referring to the institution where the mentor-mentee relationship took place). Here, research institutions are stratified into four equal-sized tiers based on the total number of citations received by all papers published by such institutions after the year 2000. The error bars in (\textbf{b}) and (\textbf{c}) represent the 95\% confidence intervals obtained via bootstrap resampling 5000 times with replacement. The results obtained via bootstrap testing for the null hypothesis of equal means between adjacent bars in the histograms are reported in (\textbf{c}) on top of the histograms. ***$p < 0.01$, **$p < 0.05$, *$p < 0.1$.}
    \label{fig:Mobility}
\end{figure*}

\begin{figure*}[ht!]
\centering
    \includegraphics[width=13cm]{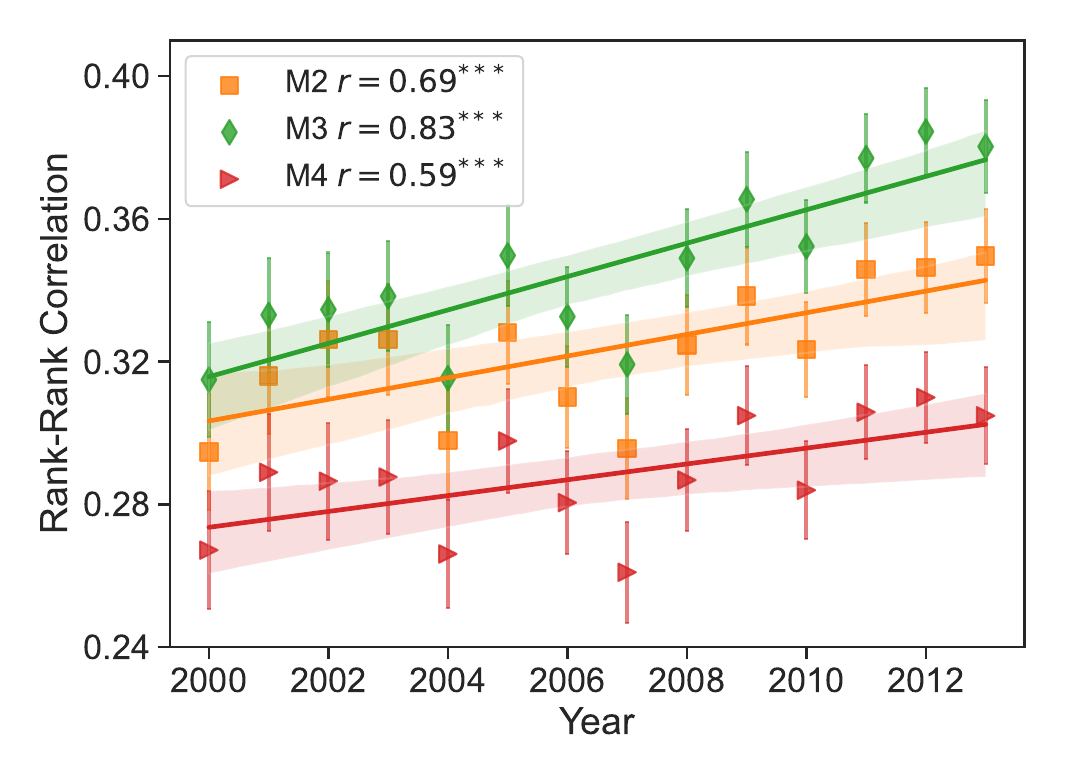}
    \caption{\textbf{Trends in mentor-mentee impact mobility under different measures of scientific impact.} Like in Fig.~\ref{fig:Mobility}B, the impact mobility here is measured as the Spearman's rank correlation coefficient for the association between the impact rankings of mentors and mentees, using data on cohorts of mentor-mentee pairs with the same final
    mentorship year. In M2, we consider the impact of a mentor/mentee by the normalized citations adjusted over time and disciplines in MAG dataset instead of raw citations used in Fig.~\ref{fig:Mobility}B. To gauge the reputation of mentors accrued over time, in M3, we have expanded the time window to measure the scientific impact of mentors, i.e., accumulating the raw citations of mentors from the year of their first publication recorded in MAG dataset (after 1980) to the year of mentorship stopped. The measure of mentee impact remains the same as in M1. In M4, we have, similar to M3, expanded the time window measuring the impact of mentors, and, at the same time used the normalised citations. The error bars represent the 95\% confidence intervals by bootstrap resampling 5000 times with replacement. The solid line and the shaded area represent the regression line (with annotated Pearson’s $r$ and $p$ values) and the $95\%$ confidence intervals, respectively. ***$p < 0.01$, **$p < 0.05$, *$p < 0.1$.}
    \label{fig:Evolution_method}
\end{figure*}

\begin{figure*}[ht!]
\centering
    \includegraphics[width=13cm]{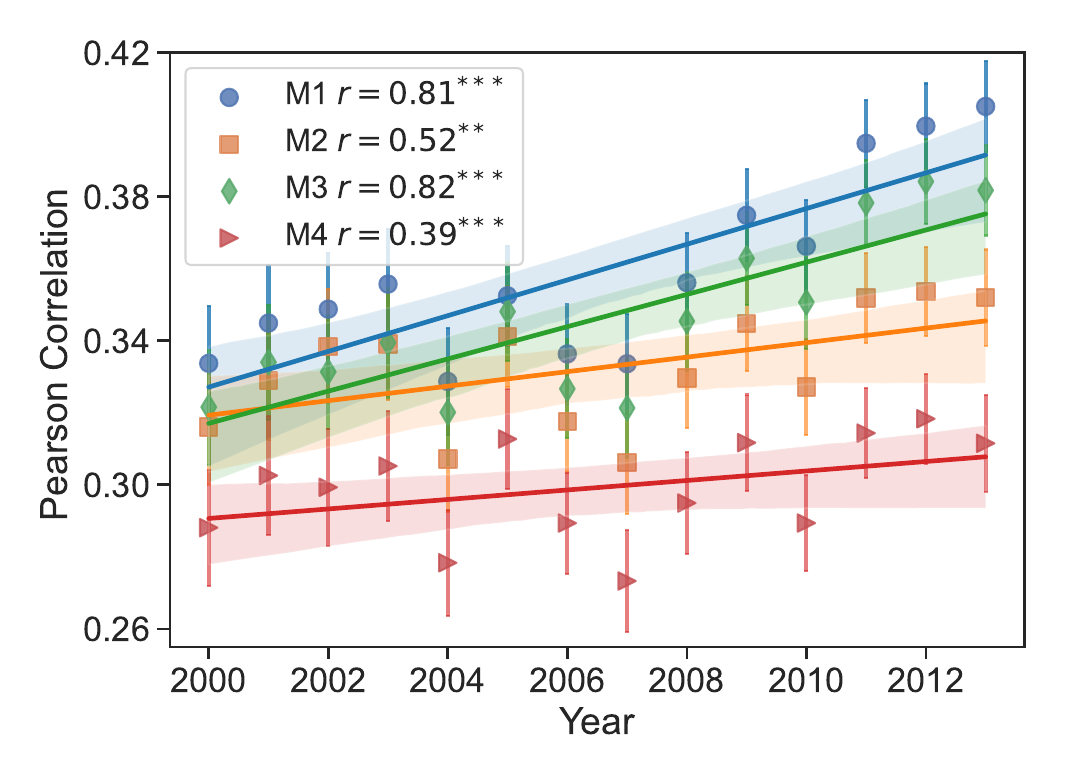}
    \caption{\textbf{Trends in mentor-mentee impact persistence measured by Pearson correlation under different measures of scientific impact.} Here, persistence is gauged as the Pearson correlation coefficient between the log impact of mentors and mentees, using data on cohorts of mentor-mentee pairs with the same final mentorship year. As in Fig.~\ref{fig:Mobility}B of the main paper, in line M1 impact of a mentor/mentee is defined as the total number of raw citations received within a 5-year time window before/after the final mentorship year. In M2, we quantify the impact of a mentor/mentee as the normalized citations adjusted over time and disciplines in the MAG dataset. To better gauge the reputation of mentors accrued over time, in M3 we have expanded the time window used to measure the impact of mentors, i.e., considering the raw citations of mentors accrued from the year of their first publication recorded in the MAG dataset (after 1980) to the final mentorship year. The measure of mentee impact remains the same as in M1. In M4, we have, similar to M3, expanded the time window measuring the impact of mentors, and, at the same time, used normalised citations. The error bars represent the 95\% confidence intervals obtained by bootstrap resampling 5000 times with replacement. The solid line and the shaded area represent the regression line (with annotated Pearson’s $r$ and $p$ values) and the $95\%$ confidence intervals, respectively. ***$p < 0.01$, **$p < 0.05$, *$p < 0.1$.}
    \label{fig:Evolution_pearson}
\end{figure*}

Fig.~\ref{fig:Mobility}a illustrates how we quantify the academic impact of mentors and mentees within a 5-year time window before and after the final year of their mentor-mentee relationship (hereafter referred to as ``final mentorship year''), such as, e.g., the year of the mentee's doctoral graduation. The aggregated impact of a mentor or mentee over this 5-year period is calculated as the sum of the citations received by their papers (within 5 years of their publication) published during such period. To analyze the persistence of impact across academic generations, we calculate the Spearman rank correlation coefficients between the impact percentile ranks of mentors and mentees for cohorts with different final mentorship years. In other words, we measure the similarity between the positions of mentors and their mentees in the impact rankings of their discipline: The higher the rank-rank correlation, the more a mentee's scientific impact is correlated to that of their mentor, the higher the intergenerational persistence. Fig.~\ref{fig:Mobility}b shows a significant upward trend in rank-rank correlations, indicating an increasing trajectory of persistence across subsequent mentor-mentee cohorts. This suggests that over time mentees have become increasingly likely to share a similar positions in their discipline's impact ranking as their own mentors. This finding is consistent with the observation that the academic impact of early-career researchers is increasingly influenced by the prominence and reputation of supervisors~\cite{ma2020mentorship} and/or collaborators~\cite{li2019early}, as well as with the existence of a ``chaperone effect'' in scientific publishing~\cite{sekara2018chaperone}. To verify that this trend is not just an artifact due to our definition of academic impact, we reevaluate impact after normalizing citations over time and disciplines, and reevaluate the impact of mentors over longer periods of time, i.e., from the year of their initial publication to the final mentorship year, reaching the same conclusion (see Fig.~\ref{fig:Evolution_method}). To further test the robustness of our results, we also measure the Pearson correlation between the logarithmic impact of mentors and mentees, once again reaching the same conclusions (see Fig.~\ref{fig:Evolution_pearson}). 

We now proceed to investigate the disparities in impact persistence across various dimensions. We compare persistence across five different mentorship types (Table.~\ref{tab:MentorshipTypeDef} and Fig.~\ref{fig:Mobility}c left), finding that \emph{Research assistants} and \emph{Collaborators} display the lowest rank-rank correlations with their mentors, followed by \emph{Research scientists} and \emph{Postdoctoral fellows}. The highest rank-rank correlation (i.e., the highest impact persistence) is observed for \emph{Graduate students}, in line with the expectation that supervisors may have a closer and more supportive relationship with their students, and therefore a stronger influence on their career prospects~\cite{gardner2009conceptualizing,ma2020mentorship}. In addition, we investigate impact persistence between different mentor-mentee gender combinations in the middle panel of Fig.~\ref{fig:Mobility}c. Our results show a slightly higher persistence associated with female mentors. This is possibly due to female mentors having a lasting positive impacts on mentees~\cite{wu2022female}, or providing career development facilitation to a larger extent than male mentors~\cite{fowler2007relationship}. After controlling for the mentor's gender, we find no statistically significant difference in persistence among mentees of different genders. Furthermore, to understand whether intergenerational impact persistence varies according to the prestige of institutions in which the mentorship took place, we first rank institutions based on the total number of citations received by papers published by authors affiliated with them (as a proxy of their prestige), and then divide institutions equally into four tiers based on the quartiles of such ranking. The right panel of Fig.~\ref{fig:Mobility}c reveals that the impact persistence between mentors and mentees is negatively correlated with the prestige of their institution, implying that persistence is relatively lower for top-tier institutions.

\begin{figure*}[ht!]
\centering
    \includegraphics[width=16cm]{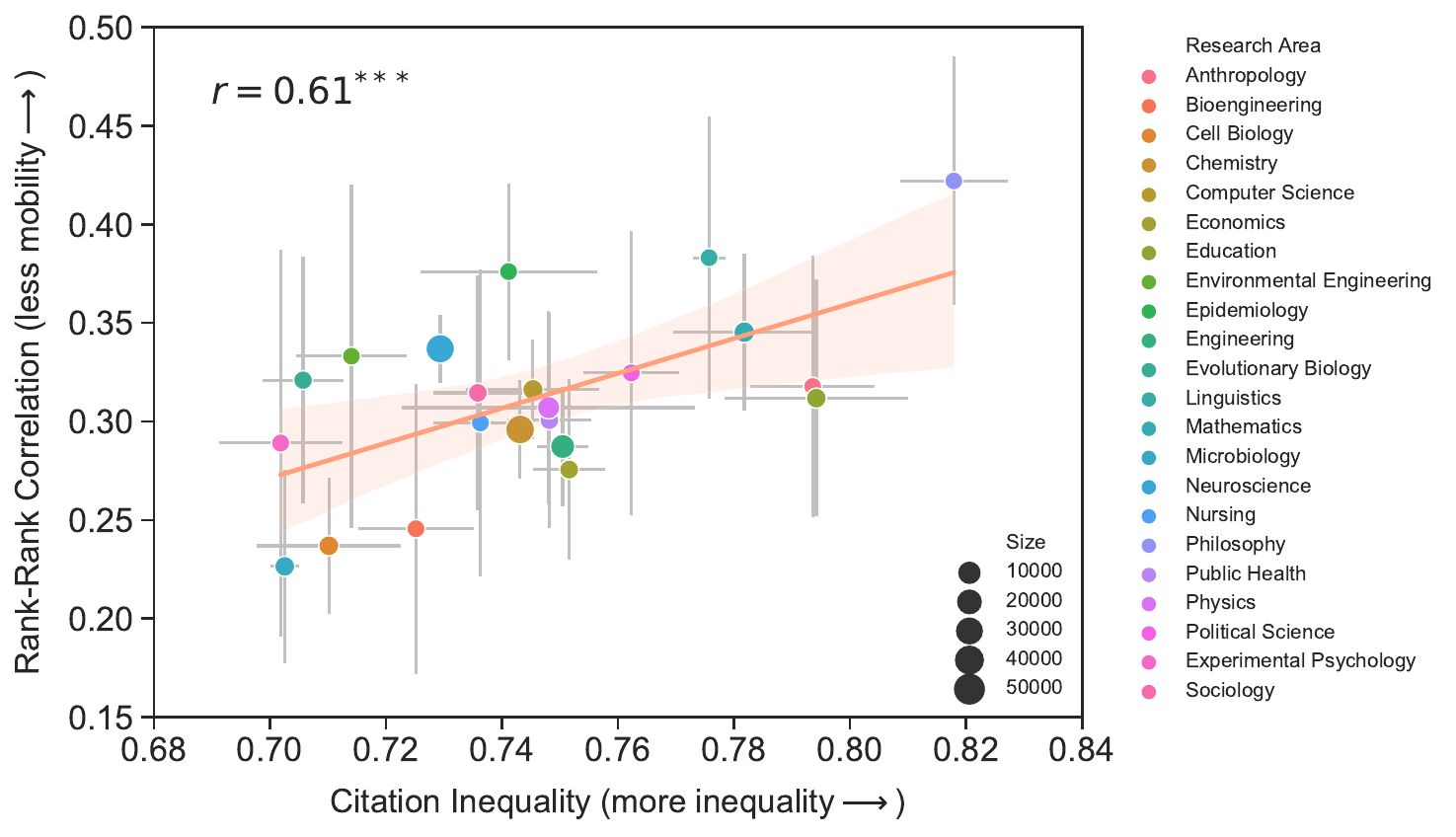}
    \caption{\textbf{The academic Great Gatsby Curve: More inequality is associated with more impact persistence across academic generations.} Citation inequality is measured by the Gini coefficient, using the cumulative number of citations authors have received from the papers published within a 5-year time window before the final mentorship year, within 5 years after publication. The error bars in vertical and horizontal directions refer to the standard deviation of the mean over those years. The point size of each research area is proportional to the number of mentor-mentee pairs considered in our analysis. The solid line and the shaded area represent the regression line (with annotated Pearson’s $r$ and $p$ values) and the $95\%$ confidence intervals, respectively. ***$p < 0.01$, **$p < 0.05$, *$p < 0.1$.}
    \label{fig:Correlation}
\end{figure*}

\begin{figure*}[ht!]
\centering
    \includegraphics[width=16cm]{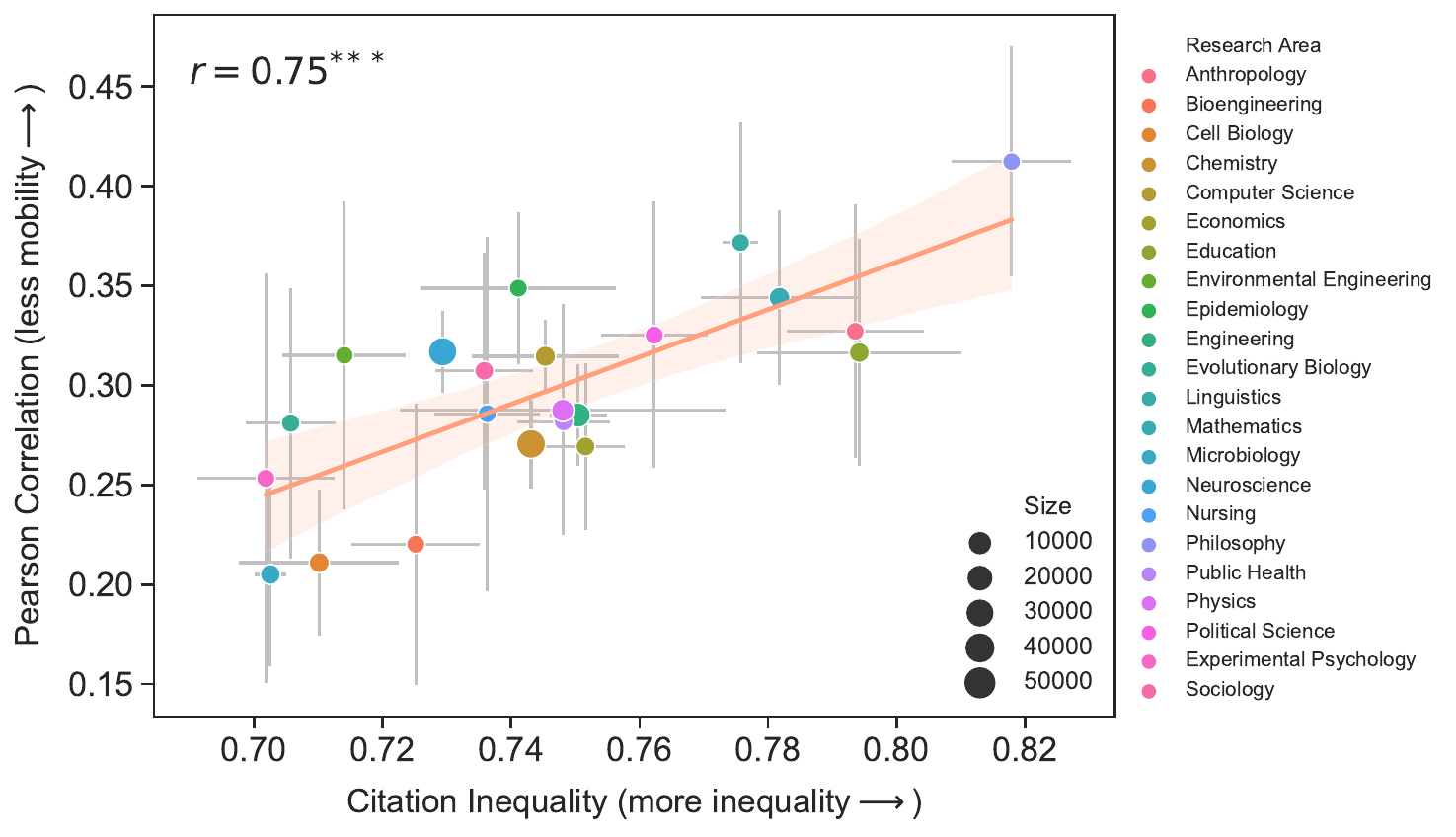}
    \caption{\textbf{Relationship between mentor-mentee impact persistence and citation inequality by research area.} Here, impact persistence is gauged as the Pearson correlation coefficient  between the log impact of mentors and mentees, using data on cohorts of mentor-mentee pairs with the same final mentorship year. Citation inequality is measured as the Gini coefficient, using the cumulative number of citations authors have received from papers published within a 5-year time window before the final mentorship years, in the 5 years after publication. The error bars refer to the standard deviation of the mean over those years. The point size of each research area is proportional to the number of mentor-mentee pairs considered in our analysis. The solid line and the shaded area represent the regression line (with annotated Pearson’s $r$ and $p$ values) and the $95\%$ confidence intervals, respectively. ***$p < 0.01$, **$p < 0.05$, *$p < 0.1$.}
    \label{fig:correlation_pearson}
\end{figure*}

\begin{figure*}[ht!]
\centering
    \includegraphics[width=18cm]{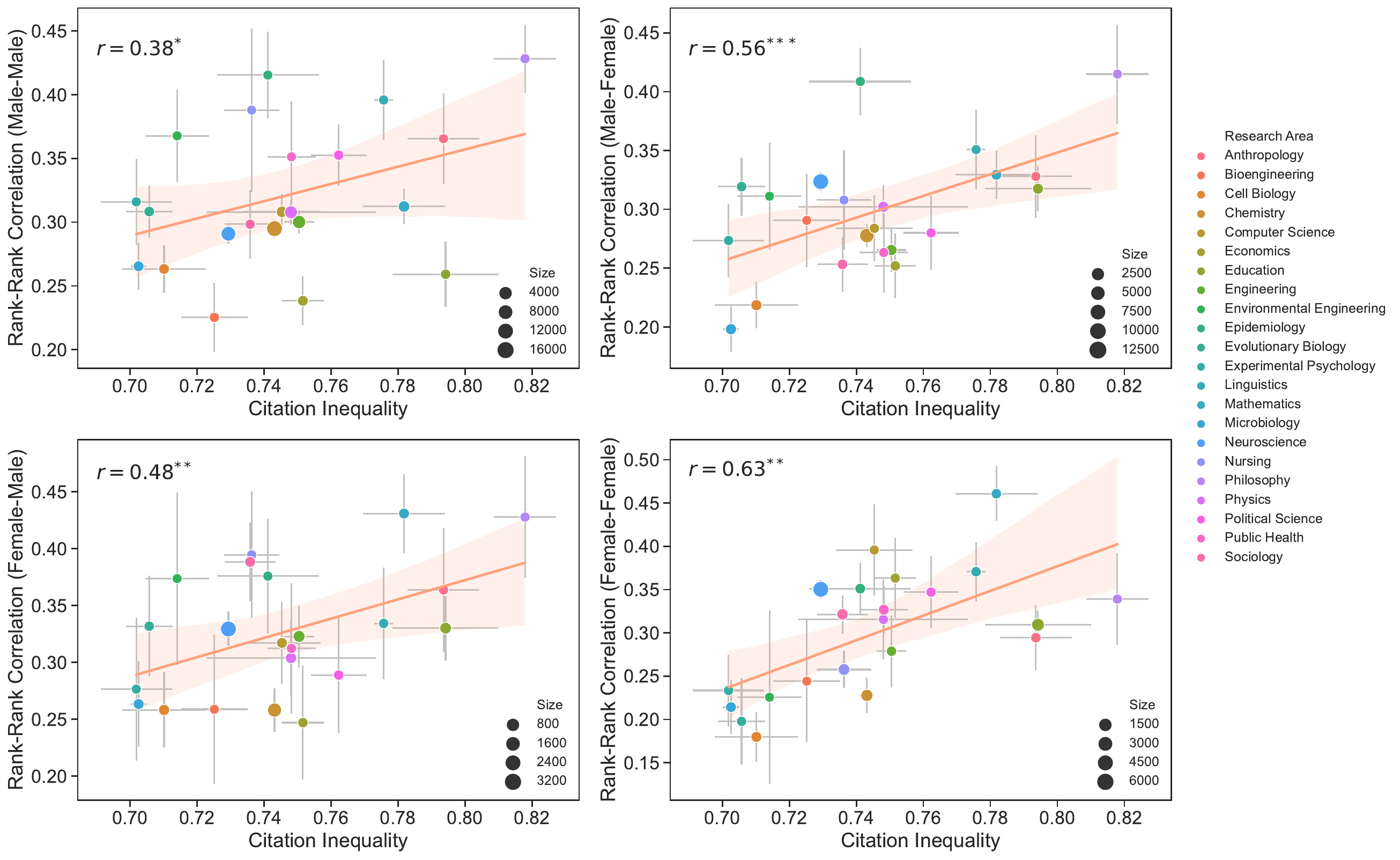}
    \caption{\textbf{The academic Great Gatsby Curve across different mentor-mentee gender combinations.} Citation inequality is measured with the Gini coefficient, using the cumulative number of citations researchers have received from the papers published within a 5-year time window before the final mentorhsip year, in the 5 years after publication. The error bars refer to the standard deviation of the mean over those years. The point size of each research area is proportional to the number of mentor-mentee pairs considered in our analysis. The solid lines and the shaded areas represent the regression lines (with annotated Pearson’s $r$ and $p$ values) and the $95\%$ confidence intervals, respectively. ***$p < 0.01$, **$p < 0.05$, *$p < 0.1$.}
    \label{fig:correlation_gender}
\end{figure*}

\begin{figure*}[ht!]
\centering
    \includegraphics[width=18cm]{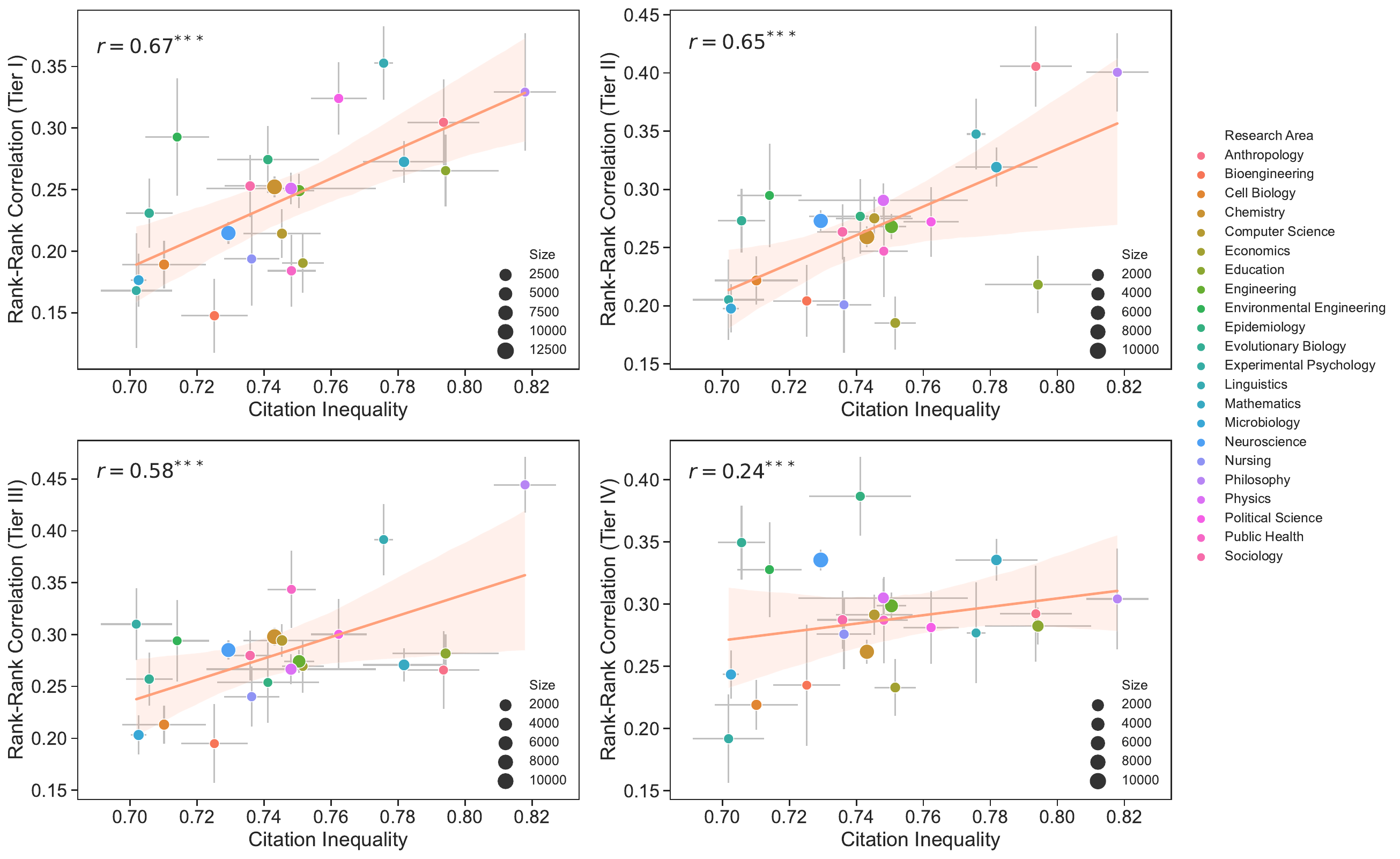}
    \caption{\textbf{The academic Great Gatsby Curve across different institutional prestige tiers.} Citation inequality is measured as the Gini coefficient, using the cumulative number of citations researchers have received from the papers published within a 5-year time window before the final mentorhsip year, in the 5 years after publication. The error bars refer to the standard deviation of the mean over those years. The point size of each research area is proportional to the number of mentor-mentee pairs considered in our analysis. The solid lines and the shaded areas represent the regression lines (with annotated Pearson’s $r$ and $p$ values) and the $95\%$ confidence intervals, respectively. ***$p < 0.01$, **$p < 0.05$, *$p < 0.1$.}
    \label{fig:correlation_institute}
\end{figure*}

Inspired by the Great Gatsby Curve in the Social Sciences~\cite{corak2013income}, we examine the association between impact inequality and intergenerational impact persistence. Fig.~\ref{fig:Correlation} ranks the disciplines included in our analysis along these two dimensions. The horizontal axis shows the impact inequality in a research area, measured by the Gini coefficient of the distribution of citations received by its authors. Over the past decade, researchers in areas such as \emph{Experimental Psychology}, \emph{Microbiology} and \emph{Evolutionary Biology} experienced the most egalitarian citation distribution, while those in \emph{Philosophy}, \emph{Education} and \emph{Anthropology} the most unequal. The vertical axis shows intergenerational impact persistence, obtained via the \emph{Spearman} rank correlation as explained above. In disciplines like \emph{Microbiology}, \emph{Cell Biology} and \emph{Bioengineering}, the correlation between the impact of mentors and mentees is the weakest. On the contrary, in some liberal arts and medical disciplines like \emph{Philosophy}, \emph{Linguistics} and \emph{Epidemiology} the persistence of impact across academic generations is rather strong. The values of mentor-mentee impact correlation and inequality in each specific research area are summarized in Table~\ref{tab:AreaMobIne}. More importantly, we observe a significant positive correlation between these two quantities (with Pearson's $r=0.61$, $p < 0.01$), suggesting that disciplines with greater inequality in their distributions of impact also tend to be areas in which academic impact is more likely to be passed on from mentors to mentees. We ran robustness checks using Pearson correlation instead of Spearman to quantify persistence, finding a consistent positive correlation between impact persistence and inequality (Fig.~\ref{fig:correlation_pearson}), with Pearson's $r=0.75$, $p < 0.01$. The positive correlation between impact persistence and inequality remains even when controlling for different mentor-mentee gender combinations (Fig.~\ref{fig:correlation_gender}) and levels of institutional prestige (Fig.~\ref{fig:correlation_institute}). Taken together, these results suggest a robust negative association between impact inequality in a given discipline and the potential for its early-career researchers (i.e., mentees) to achieve upward mobility.


\section*{Discussion}
In this paper, we find that academia is not immune from the phenomenon of intergenerational persistence, which has been widely documented in the Social Sciences across dimensions such as income, wealth and occupation~\cite{chetty2014land,xie2022trends}. We examined intergenerational academic persistence by analogizing academic mentors and mentees to parents and children, and academic impact (as measured with citations) to income. The persistence of income through genealogical generations and the persistence of impact through academic ones both reflect the transmission of resources and status, and they capture the extent to which the success of one generation may depend on that of the previous one. However, while there is a clear analogy between the mechanisms at play in these two contexts, there are also obvious differences. On one hand, both mechanisms involve the inheritance of a network, be it social, professional, or both~\cite{robison2002social,morgan2022socioeconomic}. On the other hand, the transfer of economic status is --- to a good extent --- mechanistic, as it is grounded upon the inheritance of wealth. The transfer of academic status is instead grounded upon the inheritance of intangibles~\cite{lienard2018intellectual}, such as knowledge and visibility. 

Our findings suggest that academia has become less open and more stratified over time, as newer prot\'eg\'e cohorts are characterized by lower intergenerational mobility than their predecessors. We also demonstrated that there are significant differences in impact persistence across different types of mentorship, mentor-mentee gender combinations, and levels of institutional prestige.

Finally, we demonstrated the existence of an ``academic Great Gatsby Curve'', i.e., of a positive relationship between academic impact inequality and intergenerational persistence, in analogy with the Great Gatsby Curve observed between income inequality and intergenerational persistence in the Social Sciences. This result makes it clear that academic impact --- as quantified by citations --- is to some extent inherited. As such, citation-based bibliometric indicators should be handled with care when used to assess the performance of academics.

\section*{Methods}
\subsection*{Dataset} We collected genealogical data on mentorship relationships from the Academic Family Tree (AFT, \href{https://academictree.org}{Academictree.org}), including 245,506 mentor-mentee relationships among 304,395 authors who published 9,809,145 papers across 22 disciplines. For each author, we record the person's ID, name, gender, affiliation and discipline. For each mentor-mentee relationship, we record the IDs of the mentor and mentee, the mentorship type (i.e., graduate student, postdoc or research assistant), the institution where the mentorship took place, and the first and final mentorship years. Our analysis is based on mentorship relationships that ended between 2000 and 2013. The reason we use 2000 as the starting point of our analysis is that before that year records of mentor-mentee pairs in our data are much sparser and fluctuate significantly from year to year. We use 2013 as the final year to keep track of publications for a period of five years after the final mentorship year, plus an additional five years to allow for the accumulation of citations received by such publications.

We merged the aforementioned genealogy data with the authors' publication records, citations and institutional affiliations by linking AFT with the Microsoft Academic Graph (MAG), one of the largest multidisciplinary bibliographic databases. One advantage of using the MAG database is that all entities in it (i.e., scientists, institutions and publications) have already been disambiguated and associated with unique identifiers, allowing for a sequential matching between AFT and MAG authors and affiliations. The integrated AFT and MAG data have been obtained from Ref.~\cite{ke2022dataset}, and citation information of publications authored by AFT authors is retrieved from the MAG database.  

\subsection*{Measures of mentor-mentee impact persistence}

\textbf{Spearman rank correlation.} As one of the most common measures to of intergenerational persistence, it quantifies the extent to which a mentee's impact rank tends to be associated with their mentor's impact rank, without requiring the relationship between the two to be log-linear. This measure provides a concise summary of positional persistence, which is independent of any changes in the distribution of citations between the two generations. Therefore, it can  be easily used to make comparisons across disciplines for temporal analyses.
\\
\noindent\textbf{Pearson correlation.} An alternative measure of intergenerational persistence is the Pearson correlation, which captures the correlation between the logarithmic citation impact of mentors and mentees:
\begin{equation}
    r = \frac{Cov(M_{t-4,t},P_{t,t+4})}{\sqrt{Var(M_{t-4,t})Var(P_{t,t+4})}}
\end{equation}
where $M_{t-4,t}$ and $P_{t,t+4}$ are, respectively, the logarithmic citation impact of mentors and mentees received within a 5-y time window before and after the final mentorship year $t$.

\bibliographystyle{naturemag}
\bibliography{scibib}

\section*{Data availability}
The data used in the study are publicly available from the Academic Family Tree (\href{https://academictree.org}{https://academictree.org}) and the Microsoft Academic Graph (\href{https://zenodo.org/record/6511057}{https://zenodo.org/record/2628216}). The link between the two data is derived from \href{https://zenodo.org/record/4917086}{https://zenodo.org/record/4917086}. All other data are included in the manuscript and/or SI Appendix.


\section*{Acknowledgements}
Y.S. and G.L. acknowledge support from a Leverhulme Trust research project grant (RPG-2021-282).

\section*{Author contributions}
Y.S., F.C. and G.L. conceived and designed research; Y.S. and X.L. collected the data; Y.S. performed research; Y.S., F.C., X.L. and G.L. analyzed data; Y.S., F.C., X.L. and G.L. wrote and edited the paper.

\section*{Competing interests}
The authors declare that they have no competing interests.

\clearpage

\beginsupplement

\begin{titlepage}
    \begin{center}
        \huge
        Supplementary Information for:\\
        \huge
         The academic Great Gatsby Curve\\
        \vspace{0.5cm}
        \Large
        Ye Sun$^{1}$, Fabio Caccioli$^{1,2,3}$, Xiancheng Li$^{4}$, Giacomo Livan$^{5,1,\ast}$\\
        
        \vspace{0.5cm}
        \large
        $^{1}$Department of Computer Science, University College London, 66-72 Gower Street, London WC1E 6EA, United Kingdom\\
        $^{2}$London School of Economics and Political Science, Systemic Risk Centre, London WC2A 2AE, United Kingdom\\
        $^{3}$London Mathematical Laboratory, United Kingdom\\
        $^{4}$School of Business and Management, Queen Mary University of London, Mile End Road, London E1 4NS, United Kingdom\\
        $^{5}$Dipartimento di Fisica, Universit\`a degli Studi di Pavia, via Bassi 6, 27100 Pavia, Italy
        \vspace{0.5cm}
        \large
        $^{\ast}$Corresponding author. Email: giacomo.livan@unipv.it
        \end{center}
\end{titlepage}


\clearpage

\begin{table*}
\caption{Mentorship type definitions and statistics. Persistence is quantified as the Spearman rank-rank correlation between the percentile positions of mentors and mentees in their discipline's impact rankings.}
\resizebox{\textwidth}{!}{\begin{tabular}{llcc}
\hline
\hline
Mentorship type & Description & Count &  Persistence (Spearman)  \\
\midrule
1. Research assistant    & Undergraduate, pre-bachelor's degree  & 4,444  & 0.16\\
2. Graduate student      & Work lead to master's or doctoral dissertation & 231,431  & 0.36 \\
3. Postdoctoral fellow   & Short-term employment after earning doctorate & 11,909 & 0.27 \\
4. Research scientist    & Long-term employment after doctorate & 846 & 0.29\\
5. Collaborator          & Non-directional, work together influenced each other's thinking & 1,077 & 0.17\\
\bottomrule
\end{tabular}}
\label{tab:MentorshipTypeDef}
\end{table*}

\begin{table*}
\caption{Impact rank persistence and inequality by discipline. Persistence is quantified both with the Spearman rank-rank correlation between the percentile positions of mentors and mentees in their discipline's impact rankings, and with the Pearson's correlation between their impact as quantified in terms of citations accrued by their publications.}
\resizebox{\textwidth}{!}{\begin{tabular}{llccc}
\hline
\hline
Research area & Size & Persistence (Spearman) & Persistence (Pearson) & Inequality \\
\midrule
1. Anthropology & 2,686 & 0.32 & 0.33 & 0.79 \\
2. Bioengineering & 3,366 & 0.25 & 0.22 & 0.73 \\
3. Cell Biology & 10,104 & 0.24 & 0.21 & 0.71 \\
4. Chemistry & 50,892 & 0.30 & 0.27 & 0.74 \\
5. Computer Science & 11,864 & 0.32 & 0.31 & 0.75 \\
6. Economics & 6,795 & 0.28 & 0.27 & 0.75 \\
7. Education & 9,758 & 0.31 & 0.32 & 0.79 \\
8. Environmental Engineering & 2,798 & 0.33 & 0.31 & 0.71 \\
9. Epidemiology & 3,437 & 0.38 & 0.35 & 0.74 \\
10. Engineering & 28,396 & 0.29 & 0.28 & 0.75 \\
11. Evolutionary Biology & 5,120 & 0.32 & 0.28 & 0.71 \\
12. Linguistics & 3,232 & 0.38 & 0.37 & 0.78 \\
13. Mathematics & 13,359 & 0.35 & 0.34 & 0.78 \\
14. Microbiology & 9,644 & 0.23 & 0.21 & 0.70 \\
15. Neuroscience & 49,028 & 0.34 & 0.32 & 0.73 \\
16. Nursing & 3,509 & 0.30 & 0.29 & 0.74 \\
17. Philosophy & 2,628 & 0.42 & 0.41 & 0.82 \\
18. Public Health & 5,483 & 0.30 & 0.28 & 0.75 \\
19. Physics & 19,761 & 0.31 & 0.29 & 0.75 \\
20. Political Science & 3,815 & 0.32 & 0.33 & 0.76 \\
21. Experimental Psychology & 3,891 & 0.29 & 0.25 & 0.70 \\
22. Sociology & 6,319 & 0.31 & 0.31 & 0.74\\
\bottomrule
\end{tabular}}
\label{tab:AreaMobIne}
\end{table*}

\clearpage

\end{document}